\documentclass[useAMS,usenatbib]{mn2e}
\usepackage{graphicx}
\title[An Arecibo Drift-Scan Survey]
{Discovery of 10 pulsars in an Arecibo drift-scan survey}
\author[D.~R. Lorimer et al.]
{D.~R. Lorimer$^1$,\thanks{Email: Duncan.Lorimer@manchester.ac.uk}
K.~M. Xilouris$^2$,
A.~S. Fruchter$^3$,
I.~H. Stairs$^4$,
F. Camilo$^5$,
\newauthor
A.~M. Vazquez$^6$,
J.~A. Eder$^6$,
M.~A. McLaughlin$^1$,
M.~S.~E. Roberts$^7$,
\newauthor
J.~W.~T. Hessels$^7$ and
S.~M. Ransom$^8$ 
\newauthor
\\
$^1$ University of Manchester,
Jodrell Bank Observatory, Macclesfield, Cheshire, SK11~9DL\\
$^2$ University of Virginia, Department of Astronomy, PO Box 3818,
Charlottesville, VA 22903, USA\\
$^3$ Space Telescope Science Institute, 3700 San Martin Drive, Baltimore, 
MD 21218, USA\\
$^4$ Department of Physics and Astronomy, University of British Columbia,
6224 Agricultural Road, Vancouver, BC V6T 1Z1, Canada\\
$^5$ Columbia Astrophysics Laboratory, Columbia University, 550 West 120th Street, New York, NY 10027, USA\\
$^6$ Arecibo Observatory, HC3 Box 53995, Arecibo, PR 00612, USA\\
$^7$ Physics Department, McGill University, 3600 University Street, Montreal, QC H3A 2T8, Canada\\
$^8$ National Radio Astronomy Observatory, 520 Edgemont Road, Charlottesville, VA 22903, USA
}
\begin{document}
\date{Accepted 2005 March 8; Received 2005 March 7; in original form
2005 Februrary 24}
\def\lapp{\ifmmode\stackrel{<}{_{\sim}}\else$\stackrel{<}{_{\sim}}$\fi}
\def\gapp{\ifmmode\stackrel{>}{_{\sim}}\else$\stackrel{>}{_{\sim}}$\fi}
\maketitle

\label{firstpage}

\begin{abstract}
We present the results of a 430-MHz survey for pulsars conducted
during the upgrade to the 305-m Arecibo radio telescope. Our survey
covered a total of 1147~deg$^2$ of sky using a drift-scan technique. We
detected 33 pulsars, 10 of which were not known prior to the survey
observations. The highlight of the new discoveries is PSR~J0407+1607,
which has a spin period of 25.7 ms, a characteristic age of 1.5~Gyr
and is in a 1.8-yr orbit about a low-mass ($>0.2$ M$_{\odot}$)
companion. The long orbital period and small eccentricity ($e =
0.0009$) make the binary system an important new addition to the
ensemble of binary pulsars suitable to test for violations of the
strong equivalence principle. We also report on our initially
unsuccessful attempts to detect optically the companion to J0407+1607
which imply that its absolute visual magnitude is $> 12.1$.  If, as
expected on evolutionary grounds, the companion is an He white dwarf,
our non-detection imples a cooling age of least 1~Gyr.
\end{abstract}

\section{Introduction}
\label{sec:intro}

Recycled pulsars --- neutron stars that are thought to have been spun
up by the accretion of matter from a binary companion --- are
interesting objects to study. In addition to their application as
excellent clocks for high-precision timing experiments (Davis et
al.~1985), \nocite{dtwb85} progress in our understanding of their
origin and evolution continues to be made by large-area surveys of the
radio sky.

Following Wolszczan's (1991)\nocite{wol91a} discovery of two recycled
pulsars at high Galactic latitudes, Johnston \& Bailes
(1991)\nocite{jb91} demonstrated that the local observable population
of these objects should be isotropic, and that low-frequency ($\sim 400$~MHz)
large-area surveys would be an excellent means of
probing this population.  This prediction was largely borne out by the
discovery of $\sim50$ recycled pulsars from a number of large-area
surveys undertaken in this frequency range by many groups around the
world (see, e.g., Camilo 1995 for a review).\nocite{cam95} More
recently, sensitive surveys at higher radio frequencies ($\sim 1.4$~GHz)
continue to discover millisecond and recycled pulsars in large numbers
at intermediate and low Galactic latitudes (Edwards et al.~2001;
Faulkner et al.~2004)\nocite{ebvb01,fsk+04} where interstellar
propagation effects limit the Galactic volume surveyed at lower
frequencies.

The 305-m Arecibo telescope has played a key role in 
low-frequency surveys. Between 1993 and 1998, the upgrade to the
telescope meant that it was often parked at a fixed azimuth for
extended periods. This provided a unique opportunity to survey the sky
as it drifts through the stationary telescope beam at the sidereal
rate.  At 430 MHz, the sky coverage for a declination $\delta$ is
$60\cos \delta$~deg$^2$ per day. As a number of groups were interested
in drift-scan surveys (Foster et al.~1995; Camilo et al.~1996; Lommen
et al.~2000; McLaughlin et al.~2002),
\nocite{fcwa95,cnst96,lzb+00,mac+02} the available sky was divided up
into five zones and the zenith angle of the telescope was changed
daily to cover each region uniformly.

In this paper, we report on an analysis of data in the regions
surveyed by the ``Space Telescope Science Institute and National
Astronomy and Ionospheric Center'' (STScI/NAIC) collaboration. A
preliminary account of this work was given by Xilouris et
al.~(2000)\nocite{xfl+00}.  In Section \ref{sec:obs}, we describe the
survey observations and data analysis procedure. The results,
follow-up timing and implications of the ten new pulsars found in the
survey, including the new binary pulsar J0407+1607, are detailed in
Sections \ref{sec:res} and \ref{sec:disc}.

\begin{table}
\caption{\label{tab:strips} Declination strips covered in the
STScI/NAIC survey. For each strip we list the central zenith angle
(ZA), gain ($G$), system temperature ($T_{\rm sys}$) of the 430-MHz
line feed, approximate limiting flux density ($S_{\rm min}$), area
of sky covered and the number of new and previously known pulsars
detected (new+old).}
\begin{tabular}{crccccc}
\hline
$\delta$~range & 
ZA & 
$G$ & 
$T_{\rm sys}$ &
$S_{\rm min}$ &
Area &
Pulsars\\
(deg)    &  
(deg)    & 
(K/Jy)& 
(K) & 
(mJy) &
(deg$^2$)  &
(new+old)  \\
\hline
1--2 &17      & 10.5  &  92       &  0.9           &  ~83 & 0+0 \\
6--7 &12      & 13.0  &  76       &  0.6           &  114 & 2+3  \\
11--12& 7     & 15.2  &  72       &  0.5           &  ~55 & 1+5  \\
16--17& 2     & 16.4  &  68       &  0.5           &  287 & 5+6  \\
21--22& 3     & 16.4  &  68       &  0.5           &  163 & 2+3  \\
26--27& 8     & 15.0  &  73       &  0.5           &  315 & 0+5  \\
31--32&13     & 12.5  &  80       &  0.7           &  116 & 0+1  \\
36--37&18     & 10.0  &  96       &  1.0           &  ~14 & 0+0  \\
\hline
\end{tabular}
\end{table}

\begin{table}
\caption{\label{tab:prev} STScI/NAIC survey detection statistics.
Listed are the 10 newly discovered pulsars (marked with asterisks), 23
previously known pulsars detected and 6 previously known pulsars which
lie within 5 arcmin of a pointing which were not detected.  For each
pulsar, we give the period ($P$), DM, 400-MHz flux density
($S_{400}$), offset ($\Delta$) from the beam centre, observed S/N and
the predicted S/N.  Blank S/N entries show the non detections.  Flux
densities marked with a $\dagger$ have been estimated from catalogued
$S_{1400}$ values assuming a spectral index of --1.6.}
\begin{tabular}{lrccccc}
\hline
\multicolumn{1}{c}{PSR} & $P$   & DM   & $S_{400}$& $\Delta$ &S/N&S/N\\
                        & (ms)  & (cm$^{-3}$~pc)&(mJy)& ($'$) &(obs)&(pre)\\
\hline
J0137+1654* & 414.7 & 27 & 1.4 & 1.5 & 16.7 & 13\\
J0329+1654* & 893.3 & 42 & 0.6 & 1.3 & 14.9 & 12\\
J0407+1607* & 25.7 & 36 & 10 & 0.8 & 48.7 & 45\\
J0609+2130* & 55.6 & 39 & 0.8 & 1.2 & 13.6 & 7\\
B0823+26 & 530.7 & 19 & 73 & 1.2 & 20.0 & 1802\\\\
B0919+06 & 430.7 & 27 & 52 & 2.8 & 46.4 & 389\\
B0940+16 & 1087.4 & 20 & 7 & 4.5 & 16.8 & 56\\
B1530+27 & 1124.8 & 15 & 13 & 1.9 & 30.4 & 157\\
B1534+12 & 37.9 & 12 & 36 & 4.2 & 46.5 & 238\\
J1549+2113* & 1262.4 & 25 & 0.9 & 1.2 & 14.9 & 23\\\\
J1652+2651 & 915.8 & 41 & 11 & 1.8 & 23.2 & 104\\
J1822+1120* & 1787.0 & 95 & 0.5 & 1.4 & 14.1 & 7\\
J1838+1650* & 1901.9 & 34 & 2.6 & 1.2 & 21.0 & 23\\
J1848+0604* & 2218.6 & 243 & 1.0 & 1.8 & 11.9 & 7\\
B1900+06 & 673.5 & 503 & 22 & 7.5 & 9.8 & 11\\\\
J1900+0634 & 389.8 & 323 & 1.7$\dagger$ & 3.8 & -- & 6\\
B1904+06 & 267.2 & 473 & 2.8 & 3.0 & -- & 8\\
J1905+0616* & 989.7 & 256 & 0.5 & 6.6 & 10.4 & 7\\
B1907+10 & 283.6 & 150 & 50 & 5.5 & 27.0 & 103\\
J1909+0616 & 755.9 & 352 & 2.4$\dagger$ & 6.7 & 7.6 & 3\\\\
B1911+11 & 600.9 & 100 & 1 & 2.1 & 12.0 & 6\\
B1913+16 & 59.0 & 169 & 4 & 1.8 & 12.7 & 22\\
B1913+167 & 1616.2 & 63 & 5 & 1.3 & 25.7 & 44\\
B1918+26 & 785.5 & 28 & 6 & 3.1 & 33.6 & 49\\
J1920+1110 & 509.8 & 182 & 2.8$\dagger$ & 3.8 & -- & 17\\\\
B1924+16 & 579.8 & 177 & 8 & 1.6 & 17.8 & 68\\
B1929+10 & 226.5 & 3 & 303 & 8.4 & 46.4 & 267\\
B1933+16 & 358.7 & 158 & 242 & 9.4 & 25.8 & 150\\
B1937+21 & 1.558 & 71 & 240 & 4.3 & 30.5 & 181\\
J1943+0609 & 446.2 & 71 & -- & 2.9 & 15.6 & --\\\\
J1946+2611 & 435.0 & 165 & 2 & 4.6 & -- & 5\\
J1951+1123 & 5094.0 & 31 & 1 & 1.8 & -- & 27\\
J1953+1149 & 851.8 & 140 & 2 & 4.4 & 11.8 & 19\\
J2002+1637 & 276.4 & 94 & 0.4 & 2.3 & 18.1 & 4\\
J2040+1657* & 865.5 & 51 & 0.6 & 1.4 & 18.4 & 7\\\\
J2156+2618 & 498.1 & 49 & 2.7 & 5.6 & 10.2 & 21\\
J2234+2114 & 1358.7 & 35 & 2.6 & 5.8 & 12.6 & 10\\
B2303+30 & 1575.8 & 50 & 24 & 8.7 & 19.0 & 56\\
B2315+21 & 1444.6 & 21 & 15 & 2.8 & 24.8 & 217\\
\hline
\end{tabular}
\end{table}

\begin{figure*}
\includegraphics[width=\textwidth,angle=0]{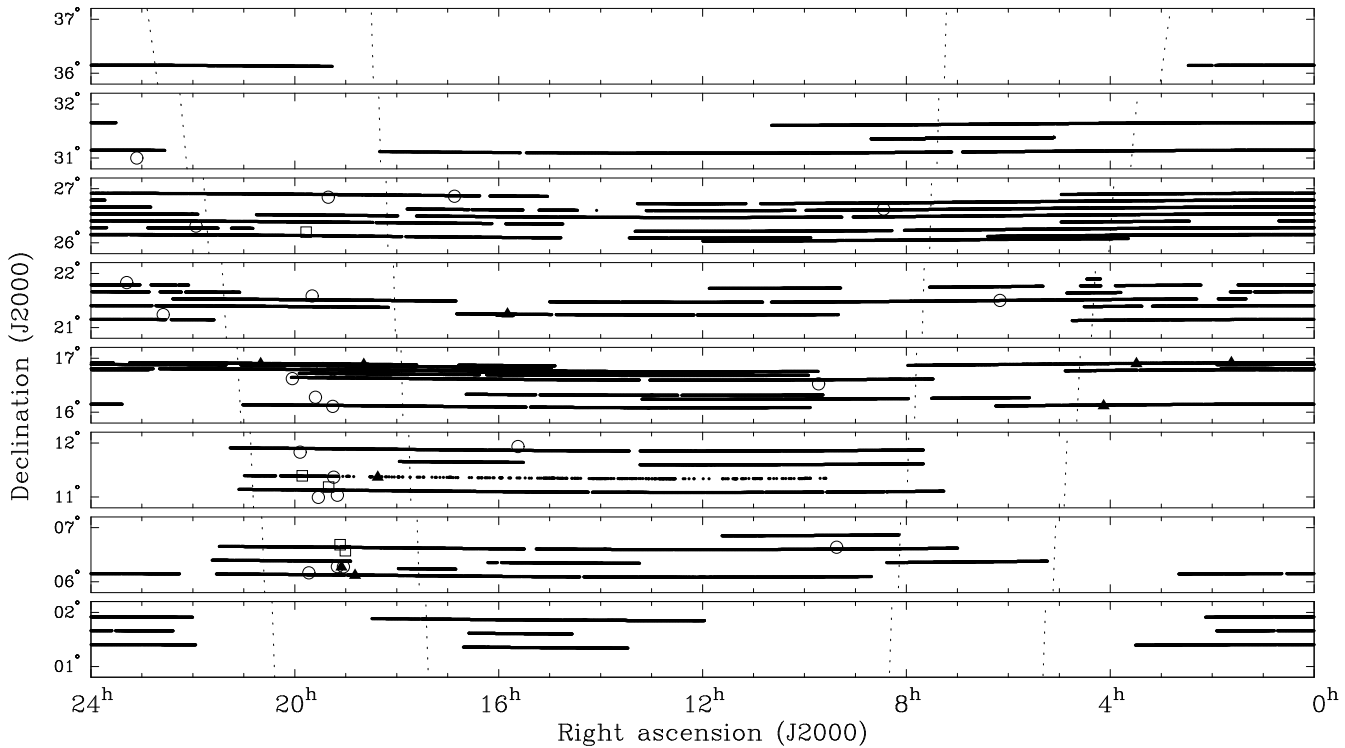}
\caption{\label{fig:sky} Survey sky coverage equatorial coordinates
showing new discoveries (filled triangles) previously known pulsars
detected by the survey (circles) and previously known pulsars not
detected (squares). Dotted lines show constant Galactic latitudes of
$\pm 20^{\circ}$.}
\end{figure*}

\section{Survey observations and results}
\label{sec:obs}

The survey observations were carried out between April 1994 and March
1995 using the 430-MHz carriage house line feed. While the nominal
sensitivity of this receiver for observations close to the zenith is
16 K~Jy$^{-1}$, these observations were made at a variety of
zenith angles and during periods when the feed was slightly out of
focus due to the maintenance work. As a result, the gain of the
telescope varied considerably in the range 10--16 K~Jy$^{-1}$.  In
drift-scan mode, a source passes through the 3~dB points of the
primary beam (10 arcmin) in about 40~s.  The declination ranges
assigned to the STScI/NAIC collaboration, area coverage and
approximate system parameters are summarised in
Table~\ref{tab:strips}.  The sky coverage is shown in
Fig.~\ref{fig:sky}. An ASCII file containing the exact sky positions
observed is available at 
http://www.blackwellpublishing.com/products/journals/
suppmat/MNR/MNR9005/MNR9005sm.htm.

The incoming radio signals were sampled by the line feed as two
orthogonal circular polarisations which, after amplification and
down-conversion, were passed to a $32 \times 250$-kHz analogue
filterbank spectrometer which sums the two polarisation channels prior
to detection with an RC time constant of 333~$\mu$s. The resulting
total-power signals were sampled every 250~$\mu$s before being
digitised with 3-bit precision and written to magnetic tape for
off-line analysis.

The data were processed in overlapping 33-s segments using a
derivative of the software package developed for earlier Arecibo
pulsar searches (Fruchter 1989; Nice 1992; Nice, Fruchter \& Taylor
1995).\nocite{fru89,nic92,nft95} In brief, the data were de-dispersed
for 448 different assumed values of dispersion measure (DM) in the
range 0--504 cm~pc$^{-3}$.  The resulting time series were then
Fourier transformed so that the amplitude spectra could be searched
for periodic signals buried in the noise. To increase sensitivity to
signals with narrow duty cycles (i.e.~many harmonics in the Fourier
domain) the spectra were incoherently summed so that up to 2, 3, 4, 8
and 16 harmonics were combined. Signal-to-noise ratios (S/Ns) were
computed from the amplitude of each point in the spectrum and the
local root-mean-square deviation. The Fourier components of the 100
strongest events were then inverse transformed back into the time
domain to compute the S/N of the resulting pulse profile. The
profiles, periods, S/Ns and best DMs of each candidate were saved for
visual inspection.

The sensitivity of the survey was dependent on the telescope's zenith
angle and hence varied for the different declination strips.  In
Table~\ref{tab:strips} we quote our theoretical minimum detectable
flux density ($S_{\rm min}$) to long-period pulsars with duty cycles
of 4\% using the radiometer equation (see, e.g., Lorimer \& Kramer
2005).  This calculation assumes a sky background temperature of 30~K
and that the pulsar transits through the beam centre.  These values
should therefore be taken as lower limits to the true survey
sensitivity which degrades for larger background temperatures, pulse
widths and shorter pulse periods.  Typical values for millisecond
pulsars, for example, are expected to be $\gapp 3$~mJy. A more
detailed discussion on the sensitivity of the observing system
can be found in Camilo, Nice \& Taylor (1996). \nocite{cnt96}

The data analysis resulted in the detection of 33 pulsars in
1147~deg$^2$ of sky. Ten of these pulsars were previously unknown
prior to the survey observations.  In Table~\ref{tab:prev}, we list
the S/Ns of the pulsars along with an estimate of the theoretical S/N
based on the nominal survey parameters and accounting for position
offsets from the centre of the (assumed Gaussian) telescope beam power
pattern. In general, for S/Ns below 20, the observed and theoretical
values are in reasonable agreement. For some of the bright pulsars
(e.g.~B0823+06), the detected S/Ns are much less than the theoretical
values.  We are unsure of the reasons for this, and, unfortunately,
the raw data are no longer available. Given our short integration
times, it is possible that the observations were affected by adverse
scintillation and/or nulling.

Also included in Table~\ref{tab:prev} are 6 previously known pulsars
that lie within 5~arcmin of the searched area but were not
detected. Although one of these pulsars, J1909+0616, made the list of
candidates with a S/N = 7.6, this was below the nominal survey
threshold S/N of 9 and does therefore not count as an blind
detection. Only two of the remaining five pulsars have theoretical
S/Ns above this threshold. For PSR~J1920+1110, we are unsure of its
430-MHz flux density which has been extrapolated from the 1400-MHz
value (Morris et al.~2002) assuming a spectral index of --1.6. As this
pulsar was not detected in more sensitive 430-MHz observations by Bhat
et al.~(2004), either the spectrum is flatter than assumed or the
pulse is significantly affected by scattering at 430~MHz. The
remaining pulsar which is nominally above our detection threshold,
J1951+1123, has a period of 5.1~s and is known to null for several
periods (Nice 1992). This is the most likely explanation for our 
non-detection.

In summary, given the uncertainties in pointing and unknown variations
in received focus during these observations, this simple analysis of
the detection statistics shows that our survey sensitivity is consistent 
with that reported by other groups (see, e.g., Camilo et al.~1996).

\begin{table*}
\caption{\label{tab:isolatedpsrs} Parameters for the nine isolated
pulsars derived from the TEMPO timing analysis.  Figures in
parentheses are 1-$\sigma$ uncertainties in the least-significant
digits as reported by TEMPO. The arrival times from which these
ephemerides are derived are freely available on-line as part of the
European Pulsar Network (EPN) database
(http://www.jb.man.ac.uk/$^\sim$pulsar/Resources/epn).  }
\begin{tabular}{lllllrcc}
\hline
PSR & R.A.~(J2000) & Dec.~(J2000) & Period, $P$ & Epoch & \multicolumn{1}{c}{$\dot{P}$} & DM & MJD range\\
    & (h:m:s) &($^\circ$:$'$:$''$) & (s) & MJD & ($\times 10^{-15}$) & ($\rm{cm^{-3}\,pc}$) & \\
\hline
J0137+1654 & 01:37:23.88(1)  & +16:54:42.1(4)  & 0.4147630265082(5) & 52225
 &  0.01223(3) & 26.6(4) &   51329--53119\\
J0329+1654 & 03:29:08.55(3)  & +16:54:02(2)    & 0.893319660649(1) & 52225
 &  0.21525(6)  & 42.1(2)   &   51329--53119\\
J0609+2130 & 06:09:58.8861(2)& +21:30:2.84(9)  & 0.05569801392951(2)& 52758
 &   0.000235(2) & 38.73(2) &   52395--53119\\ \\
J1549+2113 & 15:49:40.941(5)  & +21:13:26.9(1) & 1.262471311613(5)  & 51738
 &  0.8541(8)   & 24.8(2) &   51375--52045\\
J1822+1120 & 18:22:14.60(1)  & +11:20:56.2(2)  & 1.78703680542(4)  & 51602
 &   2.346(4)   & 95.2(6)  &   51330--51873\\
J1838+1650 & 18:38:43.062(4) & +16:50:16.05(6) & 1.90196739931(1)  & 51602
 &   2.677(1)  & 33.9(2)   &   51330--51873\\ \\
J1848+0604 & 18:48:54.622(7)  & +06:04:46.8(3) & 2.21860264699(3)  & 51602
 &   3.736(6)   & 242.7(7)  &   51330--51873\\
J1905+0616 & 19:05:06.849(5)  & +06:16:16.7(1) & 0.989706023196(7) & 51631
 & 135.232(1)  & 256.05(1)  &   51330--51931\\
J2040+1657 & 20:40:17.865(3)  & +16:57:30.46(7)& 0.865606225032(4) & 51602
 &   0.5949(6)  &  50.7(2)  &   51330--51873\\
\hline
\end{tabular}
\end{table*}

\section{Follow-up observations}
\label{sec:res}

Confirmation and follow-up observations of the new discoveries began
at end of the Arecibo upgrade in 1998. These made use of the Penn
State Pulsar Machine \nocite{cad97a} (PSPM; Cadwell 1997), a $128
\times 60$-kHz channel analogue filterbank spectrometer which records
either continuously-sampled data (``search mode'') or
synchronously-averaged pulse profiles at the predicted pulse period
(``timing mode'').

As in the original survey observations, we used the 430-MHz line-feed
receiver for the bulk of our follow-up work. Following initial
confirmation observations which verified the existence of each pulsar
in search mode, we generated a preliminary ephemeris which was
subsequently used to obtain timing-mode data.  The basic observing and
analysis procedures used were identical to those described by Lorimer,
Camilo \& Xilouris (2002).\nocite{lcx02}
\begin{figure}
\includegraphics[width=6cm,angle=270]{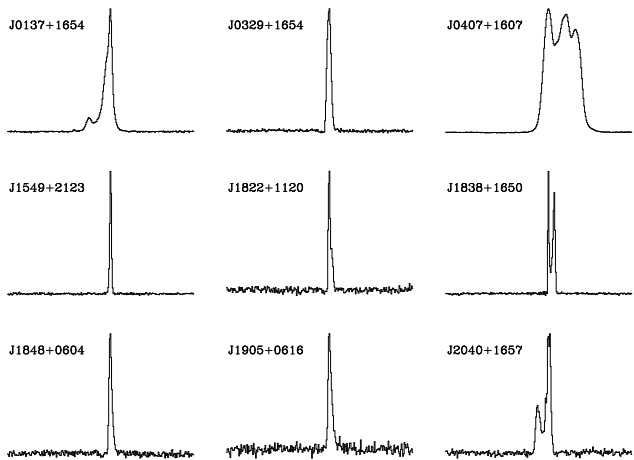}
\caption{\label{fig:profs}
Integrated 430-MHz pulse profiles for nine of the newly
discovered pulsars derived from a phase-coherent addition of the
individual PSPM observations.  Each profile shows 360 degrees of
rotational phase.  These profiles are freely available as part of the
EPN database.  A pulse profile for PSR~J0609+2130 has already been
published by Lorimer et al.~(2004) and is not included here.}
\end{figure}

In order to determine accurate spin and astrometric parameters for
each pulsar, we carried out a dedicated campaign of observations using
the PSPM in timing mode. For each observation the set of 128 folded
pulse profiles (one profile per frequency channel) was appropriately
delayed with respect to the centre observing frequency to account for
the effects of interstellar dispersion. Standard software tools
(Lorimer 2001)\nocite{lor01b} implemented this procedure to form a
de-dispersed time-tagged integrated pulse profile for each
observation.  A pulse time of arrival (TOA) for each profile was then
obtained by cross correlating it in the Fourier domain with a high S/N
template profile formed from the sum of many individual observations
(Taylor 1992).\nocite{tay92} The final template profiles are shown
in Fig.~\ref{fig:profs}.

The TOAs obtained for each pulsar were transformed to equivalent
arrival times at the solar system barycentre and fit to a simple
spin-down model using the {\sc
TEMPO}\footnote{http://pulsar.princeton.edu/tempo} software package.
After several iterations, an initial ephemeris was obtained which was
then used to form an improved template profile from the sum of each
pulse profile appropriately phase-aligned according to the timing
model. The resulting high S/N template was then used to obtain TOAs of
higher precision which were analysed with {\sc TEMPO} in a further
iteration to produce the final timing solution.  Refined values of DM
for each pulsar were obtained by forming four 1.92-MHz-wide sub-bands
from each PSPM profile. The DM was derived from a fit to the TOAs from
each sub-band keeping the spin and astrometric parameters fixed at the
final timing solution values.  For nine of the new pulsars, the TOAs
were adequately fit using an isolated pulsar spin-down model and the
resulting parameters from each fit are presented in
Table~\ref{tab:isolatedpsrs}.

In the time since the original survey observations and follow-up, five
of the new pulsars have been independently discovered in other
surveys. The binary pulsar J0407+1607 discussed below was detected
during a survey of gamma-ray error boxes at Parkes (Roberts et
al.~2004). \nocite{rrh+04} PSRs~J0609+2130 (Lorimer et
al.~2004)\nocite{lma+04} and J1549+2113 (Foster et al.~1995;
Lewandowski et al.~2004) were discovered \nocite{fcwa95,lwf+04} in
contemporary 430-MHz Arecibo drift-scan surveys.  Finally, PSRs
J1848+0604 (Faulkner, private communication) J1905+0616 (Morris et
al.~2002) were detected by the Parkes multibeam
survey. PSRs~J1549+2113 and J1848+0604 have no previously published
timing solution and are presented for the first time in
Table~\ref{tab:isolatedpsrs}. For PSR~J0609+2130, we have added an
additional year of PSPM TOAs to those already published by Lorimer et
al.~(2004). The longer time baseline of the resulting ephemeris
presented in Table~\ref{tab:isolatedpsrs} has significantly improved
the precision of the derived parameters.  \nocite{mhl+02} For
PSR~J1905+0616, we have included Jodrell Bank TOAs collected at
1412~MHz by Morris et al.~(2002) in our final timing analysis. This
has resulted in an improvement on the ephemeris published by Morris et
al.

The most interesting of the new discoveries is PSR~J0407+1607.  The
intermediate 25.7-ms spin period of this pulsar suggested that it was
either a young or recycled object.  Following initially unsuccessful
attempts to fit an isolated spin-down model to the data, sets of
closely-separated TOAs were used to obtain estimates of the
barycentric pulse period for a given epoch.  These periods were then
fit to a simple Keplerian model which revealed 1.8-yr sinusoidal
variations implying that the pulsar was a member of a long-period
circular orbit binary system.

While our early PSPM timing-mode observations could be used to obtain
a preliminary orbital ephemeris, they were not of sufficient quality
or density to obtain a phase-coherent timing solution.  In January
2002, we began a regular observing campaign to collect data with the
PSPM in search mode. These data were then folded off-line to produce a
set of pulse profiles and hence TOAs for a timing analysis. The
advantage of this approach is that the data could be re-folded later
when the precision of the orbital parameters improved.
\begin{table}
\caption{\label{tab:0407}
Measured parameters for PSR~J0407+1607 from 71 TOAs from MJDs 52304 
(January 30, 2002) to 53292 (October 14, 2004).  
The numbers in parentheses are the 1--$\sigma$ TEMPO uncertainties
in the least significant digit quoted.}
\begin{center}
\begin{tabular}{l l}
\hline
Parameter & Value \\
\hline
Right ascension (J2000) (h:m:s)                 & 04:07:54.939(3)
\\
Declination (J2000) ($^\circ$:$'$:$''$)         & 16:07:16.4(2)
\\
Spin period, $P$ (ms)                           & 25.70173919463(2)
\\
Period derivative, $\dot{P}$ ($\times 10^{-20}$)& 7.9(3)
\\
Epoch of period (MJD)                           & 52799
\\
Dispersion measure, DM (cm$^{-3}$~pc)           & 35.65(2)
\\
Projected semi-major axis, $x$ (s)              & 106.45026(2)
\\
Orbital period, $P_b$ (days)                    & 669.0704(1)
\\
Orbital eccentricity, $e$                       & 0.0009368(6)
\\
Longitude of periastron, $\omega$ (deg)         & 291.74(2)
\\
Epoch of periastron, $T_0$ (MJD)                & 52774.69(3)
\\
\hline
\end{tabular}
\end{center}
\end{table}

Despite this improved approach, small uncertainties in the assumed
position resulted in significant covariances between the orbital
parameters and position fits in TEMPO.  Fortunately, as mentioned
above, this pulsar was independently discovered by Roberts et
al.~(2004).  Follow-up 1400-MHz Arecibo observations from that survey
resulted in a positional determination which was used as a starting
point for the final timing analysis. The final phase-coherent timing
solution presented in Table \ref{tab:0407} is the result of a
simultaneous fit for spin, astrometric and binary parameters.  The
observed minus computed TOA residuals for this solution are featureless
with a root-mean-square value of 16 $\mu$s.  The DM was determined by
including 1400-MHz observations of the pulsar with the Wideband
Arecibo Pulsar Processor (Dowd, Sisk \& Hagen 2000) and, as for the
other pulsars, keeping all other model parameters constant in the
TEMPO fit.

\begin{table*}
\caption{\label{tab:derpars}
Observed and derived parameters for the ten pulsars discovered in the
survey. For each pulsar, we list the Galactic longitude ($l$) and
latitude ($b$), the flux density at 430~MHz ($S_{430}$), the pulse
width measured at 50 and 10\% of the peak intensity (respectively
$W_{50}$ and $W_{10}$), the equivalent width of a top-hat pulse having
the same area and peak intensity of the profile ($W_{\rm eq}$), the
distance derived from the dispersion measure assuming the Cordes \&
Lazio (2002) electron density model ($D$), as well as the base-10
logarithms of the 430-MHz luminosity ($L_{430}=S_{430} D^2$), the
characteristic age ($\tau=P/(2 \dot{P}) $), the surface dipole
magnetic field ($B=3.2 \times 10^{19} (P \dot{P})^{1/2}$~G) and the
rate of loss of spin-down energy ($\dot{E}=3.95 \times 10^{46}
\dot{P}/P^3$~erg~s$^{-1}$).}
\begin{tabular}{lrrlllllcccc}
\hline
\multicolumn{1}{c}{PSR} & 
\multicolumn{1}{c}{$l$} &
\multicolumn{1}{c}{$b$} &
\multicolumn{1}{c}{$S_{430}$} &
\multicolumn{1}{c}{$W_{50}$} &
\multicolumn{1}{c}{$W_{10}$} &
\multicolumn{1}{c}{$W_{\rm eq}$} &
\multicolumn{1}{c}{$D$} &
\multicolumn{1}{c}{log$_{10}$} &
\multicolumn{1}{c}{log$_{10}$} &
\multicolumn{1}{c}{log$_{10}$} &
\multicolumn{1}{c}{log$_{10}$} \\
 & 
deg &
deg &
mJy &
ms &
ms &
ms &
kpc &
$[L_{430} ({\rm mJy~kpc}^2)]$ &
$[\tau({\rm yr})]$ &
$[B({\rm G})]$ &
$[\dot{E} ({\rm erg~s}^{-1})]$\\
\hline
J0137+1654&138.3&--44.6&1.4 &14& 62& 20  & 1.2& 0.3& 8.73 & 10.9 & 30.8\\
J0329+1654&168.5&--31.7&0.6 &17& 37& 23  & 1.8& 0.3& 7.82 & 11.6 & 31.1\\ 
J0407+1607&176.5&--25.8&10.2&5.3&6.6&4.8 & 1.3& 1.3& 9.71 & 9.16 & 32.3\\
J0609+2130&189.2&1.0   &0.8 &1.6&7.4&2.4 & 1.2& 0.06& 9.57& 9.56 & 31.7\\
J1549+2113& 34.8&  49.3&0.9 &7& 22& 15  & 2.1& 0.6& 7.37 & 12.0 & 31.2\\ \\
J1822+1120& 39.8&  11.6&0.5 &25& 59& 27  & 3.8& 0.9& 7.08 & 12.3 & 31.2\\ 
J1838+1650& 46.6&  10.5&2.6 &72&92&39 & 2.0& 1.0& 7.05 & 12.4 & 31.2\\ 
J1848+0604& 37.9&   3.5&1.0 &46& 73& 40  & 7.0& 1.7& 6.97 & 12.5 & 31.1\\ 
J1905+0616& 40.1& --0.1&0.5 &21& 46& 24  & 5.7& 1.2& 5.06 & 13.1 & 33.7\\ 
J2040+1657& 61.2&--14.8&0.6 &16& 79& 31  & 3.0& 0.7& 7.36 & 11.9 & 31.6\\ 
\hline
\end{tabular}
\end{table*}
\begin{figure}
\includegraphics[width=84mm]{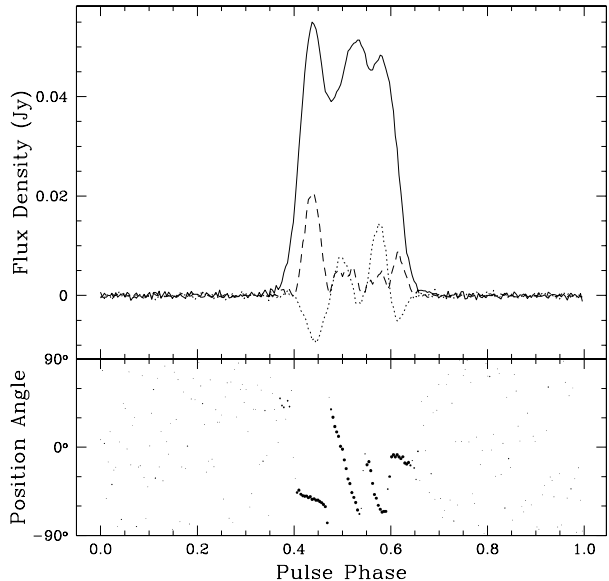}
\caption{Upper panel: full Stokes profile for PSR~J0407+1607 at
430~MHz showing total intensity (solid line), linear polarisation
(dashed) and circular polarisation (dotted) as a function of pulse
phase. Lower panel: position angle versus pulse phase.
\label{fig:pol}}
\end{figure}

To obtain an estimate of the transverse speed of J0407+1607, we
performed a scintillation analysis on an observation at 430~MHz
using the Arecibo Observatory Fourier Transform Machine (AOFTM), a
fast-dump digital spectrometer which provides 1024 frequency channels
across a 10-MHz bandwidth (see, e.g., Mclaughlin et
al.~2002).\nocite{mac+02} Performing a standard autocorrelation
analysis on the pulsed intensity as a function of time and radio
frequency (for details of this technique see Lorimer \& Kramer
2005)\nocite{lk05} we find the characteristic timescale and bandwidth
of scintillation to be $843 \pm 6$~s and $190 \pm 2$~kHz respectively.
Adopting the expressions given by Cordes \&
Rickett~(1998)\nocite{cr98} for a uniform scattering medium, and
taking a distance of $1.3\pm0.2$~kpc derived from the DM and the
Cordes \& Lazio (2002)\nocite{cl02} electron density model, we
estimate the transverse speed of PSR~J0407+1607 to be $35 \pm
5$~km~s$^{-1}$.  Although this result is model dependent, it implies
a transverse speed which lies at the low end of the distribution
known for millisecond pulsars (Nice \& Taylor 1995).  \nocite{nt95}
In future it should be possible to measure the proper motion of
this pulsar directly by extending the existing baseline of timing
observations. The predicted proper motion at this distance is $6 \pm 1$~mas
yr$^{-1}$.  Through simulations with TEMPO, we estimate that a direct
proper motion measurement of this magnitude through timing techniques
will be possible within 3~yr. The effect of proper motion on the
observed semi-major axis should also be measurable on a similar
timescale, allowing us to place constraints on the orbital inclination
angle in future.

The PSPM and AOFTM data do not provide any information about the
polarisation properties of PSR~J0407+1607. To investigate this, in
March 2002, we observed the pulsar for 78~min at 430~MHz using the
Princeton Mark IV data recording system (Stairs et
al.~2000)\nocite{sst+00} in narrow-band (5~MHz) mode. The Mark IV data
were coherently de-dispersed off-line and the data were calibrated
using observations with a noise diode of known flux density and
polarisation.  The total intensity, linear and circularly polarised
pulse profiles are shown in Fig.~\ref{fig:pol}. Relative to the
total intensity, we find that both the linear polarisation and modulus
of the circularly polarised component have an average value of 13\%,
typical of other millisecond pulsars (Xilouris et al.~1998;
Stairs, Thorsett \& Camilo~1999)\nocite{xkj+98,stc99}.  Also shown in
Fig.~\ref{fig:pol} is the position angle of the linear polarisation
as a function of pulse phase. As for many other millisecond pulsars
the position angle does not follow a smooth S-shape variation expected
from the rotating-vector model of Radhakrishnan \& Cooke
(1969).\nocite{rc69a} A determination of the emission geometry of this
pulsar is, therefore, non-trivial.

\section{Discussion}
\label{sec:disc}

Various observed and derived parameters of the new pulsars are
summarised in Table~\ref{tab:derpars} where we also include estimates
of each pulsar's 430-MHz flux density. These were derived from the
PSPM profiles using identical procedures to those described by
Lorimer et al.~(2002)\nocite{lcx02}.

\subsection{General remarks}

In common with results from other Arecibo drift-scan surveys (Foster
et al.~1995; Camilo et al.~1996; Lommen et al.~2000; McLaughlin et
al.~2002), the new pulsars are relatively faint, nearby objects. Their
inclusion in future population syntheses should provide valuable
constraints on the low end of the radio pulsar luminosity function. In
1147~deg$^2$ of sky searched, we detected five recycled pulsars:
J0407+1607, J0609+2130, B1534+12, B1913+16 and B1937+21. Of the two
new discoveries, the binary pulsar J0407+1607 is discussed below,
while the isolated pulsar J0609+2130 was already discussed in detail
by Lorimer et al.~(2004).  We simply note here that our detection rate
of one recycled pulsar per $\sim 230$~deg$^2$ is consistent with that
of other Arecibo surveys (see Camilo~1995).

The youngest pulsar in our sample, with a characteristic age of 116
kyr is PSR~J1905+0616.  As noted and discussed by Kramer et
al.~(2003), \nocite{kbm+03} this object lies within the bounds of the
EGRET source 3EG~J1903+0550.  An association is unlikely simply due to
the large density of pulsars in this region and hence high probability
of a chance aligmnent. In addition, as noted by Kramer et al., the
spin-down energy loss of J1905+0616 is not large enough to be a likely
source of the gamma-ray emission seen.

\subsection{PSR~J0407+1607 and 3EG~J0407+1610}

PSR~J0407+1607 is coincident with the unidentified EGRET source
3EG~J0407+1610 (Hartman et al.~1999),\nocite{hbb+99} a relatively weak
gamma-ray source with mean flux $F_{\gamma} = 2.4\times
10^{-10}$~erg~cm$^{-2}$~s$^{-1}$ above 100 MeV (assuming a photon
spectral index of --2). While the variability index (see McLaughlin et
al.~1996 for a definition)\nocite{mmct96} is only 0.19, and consistent
with other EGRET sources known to be associated with pulsars, our
$\dot{P}$ measurement of J0407+1607 implies a rate of loss of
spin-down energy $\dot{E}=1.9 \times 10^{32}$~erg~s$^{-1}$. This can
be used to test the validity of an association with 3EG~J0407+1610 by
calculating the required efficiency $\eta$ of conversion of the
pulsar's spin-down luminosity into gamma rays.  At the DM-derived
distance of 1.3~kpc, and assuming a gamma-ray beaming fraction of
$1/4\pi$ sr, we find an incompatible value of $\eta = (F_{\gamma}
D^2/\dot{E}) \sim 20$.  A much smaller distance and/or beaming fraction
would be required so that $\eta \lapp 1$. We therefore conclude that
PSR~J0407+1607 is most likely {\it not} the source of the gamma rays
from 3EG~J0407+1610.

\subsection{PSR~J0407+1607 and SEP violations}

The orbital parameters of the J0407+1607 binary system are consistent
with standard models of binary evolution (see, e.g., Phinney \&
Kulkarni 1994)\nocite{pk94} in which the radio pulsar has been spun up
by the accretion of matter while its companion star was in a red giant
phase. Indeed, the observed orbital eccentricity is in excellent
agreement with the prediction from Phinney's (1992)\nocite{phi92b}
relationship between eccentricity $e$ and orbital period $P_b$.

Millisecond pulsar--white dwarf binaries like PSR~J0407+1607
make excellent probes of the strong equivalence principle (SEP).
The SEP states that free fall of a body is completely independent
of its gravitational self energy.  If the SEP is violated, then the 
ratio of inertial mass to gravitational mass of a falling body differs
from unity by an amount $\Delta$. As originally proposed by Damour \&
Sch\"afer (1991)\nocite{ds91}, and by analogue with Nordvedt's earlier
tests of the SEP in the Earth-Moon system (Nordvedt
1968a,b)\nocite{nor68a,nor68b}, a violation of the SEP for a
pulsar--white dwarf binary would mean that both objects ``fall'' in the
local Galactic gravitational field at a different rate. As a result, a
violation would cause a polarisation of the orbital plane toward and
parallel to the centre of the Galaxy. 

Following Damour \& Sch\"afer's early work, Wex (1997, 2000)
\nocite{wex97,wex00} derived limits on $\Delta$ 
from an ensemble of binary pulsar-white dwarf systems. As the
figure-of-merit for this calculation is $P_b/e^2$, long-period circular orbit
binaries like J0407+1607 are particularly important members of the
ensemble used in the test. Recently, a robust Bayesian analysis by Stairs
et al.~(2005) which uses all known binary pulsars relevant for this test
including J0407+1607 find a limit on SEP violations of 
$\Delta < 5.5 \times 10^{-3}$\nocite{sfl+05}. This new limit is
comparable to the weak-field limits found in the Earth-Moon system
(Nordvedt 1968a,b).

\subsection{The companion to PSR~J0407+1607}

The observed mass function of PSR~J0407+1607 is $2.89 \times
10^{-3}$~M$_{\odot}$. Assuming a mass for the pulsar in the range
1.3--1.6~M$_{\odot}$, the implied companion mass is
0.2~M$_{\odot}$/$\sin i$, where $i$ is the unknown inclination angle
between the plane of the orbit and the plane of the sky. From
extensive studies of other millisecond pulsar binaries with similar
orbital characteristics to J0407+1607 (for a recent review, see
van Kerkwijk et al. 2005)\nocite{vbj+05}
it is most likely that the companion to J0407+1607 is a
low-mass white dwarf.  An inspection of archival Palomar data in the
digitized sky survey archive\footnote{http://www-gsss.stsci.edu/DSS}
reveals no optical counterpart down to the plate limit (approximately
19th magnitude), with the closest object in the field being 8~arcsec
($50 \sigma$) offset from the radio position given in
Table~\ref{tab:0407}.

To search for a fainter optical counterpart, the field surrounding PSR
J0407+1607 was observed on October 2003 with the 2.3-m Bok
telescope at Kitt Peak using the Steward Observatory imaging camera in
combination with the Harris B and V filters.  Four 27-min B and four
20-min V images were obtained under non-ideal seeing ($\sim 1.5$
arcsec) conditions. All images were bias-subtracted and flat-fielded
using the IRAF package\footnote{http://iraf.noao.edu}. For photometry
calibration, we selected all stars in a $3 \times 3$ arcmin box around
the pulsar location from the USNO-B1.0
Catalog\footnote{http://www.nofs.navy.mil/data/fchpix}.

The detection limit was determined to be magnitude 21 in B. This is in
good agreement with the 20.9 magnitude of the star identified as
1061$-$0042921 in the USNO-B1.0 catalogue detected at the noise level
in our images.  Four 20-min exposures were aligned and added to
produce the final V image. We used the published visual magnitude of
the nearby stars SA 94$-$171 and GD~71 (Landolt 1992)\nocite{lan92} to
determine the relationship between instrumental and aperture
magnitudes. From this procedure, we estimate a limiting apparent
visual magnitude for detection of 23.

At the nominal distance of $1.3\pm0.2$~kpc, our upper limit
corresponds to an absolute visual magnitude of at least 12.1. Assuming
the companion to be an He white dwarf, we can use the cooling curves
computed by Hansen \& Phinney (1998) \nocite{hp98a}to place a lower
limit on the age of the white dwarf companion. From an inspection of
Fig.~13 in their paper, our current non-detection constrains the
cooling age to be $\gapp 1$~Gyr, compatible with the 1.5~Gyr
characteristic age of J0407+1607.  More stringent constraints should
be available soon following deeper observations of this system now
underway with the Palomar telescope which will be described elsewhere.

\section*{Acknowledgements} 
The Arecibo observatory, a facility of the National Astronomy and
Ionosphere Center, is operated by Cornell University in a co-operative
agreement with the National Science Foundation (NSF).  We thank Alex
Wolszczan for making the PSPM freely available for use at Arecibo, as
well as Paulo Freire, Jim Sheckard, Andrea Lommen, Zaven Arzoumanian,
Ramesh Bhat and David Champion for assistance with the PSPM timing
observations. DRL is a University Research Fellow funded by the Royal
Society. IHS holds an NSERC UFA and is supported by a Discovery
Grant. FC is supported by NASA and the NSF. The
Digitized Sky Survey was produced at the Space Telescope Science
Institute under U.S. Government grant NAG W-2166. The images of these
surveys are based on photographic data obtained using the Oschin
Schmidt Telescope on Palomar Mountain and the UK Schmidt
Telescope. The plates were processed into the present compressed
digital form with the permission of these institutions.

\end{document}